# Transient Leadership and Collective Cell Movement in Early Diverged Multicellular Animals

MIRCEA R. DAVIDESCU AND IAIN D. COUZIN, Princeton University

1. INTRODUCTION: *TRICHOPLAX ADHAERENS* AS A MODEL SYSTEM FOR COLLECTIVE CELL MIGRATION

Our work tackles collective motion and decision-making in a primitive multicellular animal, *Trichoplax adhaerens*, enabling pioneering research into how intercellular coordination affects animal behavior and how migration accuracy scales with group size. Collective motion of cells is critical to some of the most vital tasks including wound healing, development, and immune response [Friedl and Gilmour 2009; Tokarski *et al.* 2012; Lee *et al.* 2012; Beltman *et al.* 2009], and is common to many pathological processes including cancer cell invasion and teratogenesis [Khalil and Friedl 2010]. The extensive understanding of movement by single cells [Rørth 2011; Insall and Machesky 2011; Houk *et al.* 2012] is insufficient to predict the behavior of cellular groups [Theveneau *et al.* 2013; Trepat, X. and Fredberg 2011], and identifying underlying rules of coordination in collective cell migration is still evasive. This is of particular interest, as collective motion and decision-making are fundamental phenomena in systems ranging from molecules [Schaller and Bausch 2013] to animal societies [Couzin *et al.* 2005], and recent studies evidence the existence of common mechanisms among disparate systems, such as group polarization by chase-and-run in neural crest cells and marching locusts [Theveneau *et al.* 2013; Bazazi *et al.* 2008]. Few of the supposed benefits of collective motion have ever been tested at the cellular scale. As an example, though collective sensing allows for larger groups to exhibit greater accuracy in navigation [Simons 2004; Berdahl *et al.* 2013] and group taxis is possible through the leadership of only a few individuals [Couzin *et al.* 2005], such effects have never been investigated in collective cell migration. It has also proven difficult to relate observed phenomena to animal behavior and fitness.

To the best of our knowledge, *T. adhaerens,* an animal of unmatched simplicity, composed of two epithelial layers, between which there exists a layer of multi-nucleate fiber cells, enables the first test of collective cell migration *in situ* outside of a developmental context. Arguably the earliest-diverged motile multicellular animal [Dellaporta *et al.* 2006], *T. adhaerens* represents not only a very important evolutionary milestone, but is also ideal for relating cellular and animal behavior. The animal has only four somatic cell types and no defined anterior-posterior asymmetry, cephalization, neurons or organs [Schierwater 2005; Ball and Miller 2010]. In spite of its apparent physical simplicity, analysis of its compact genome reveals remarkable complexity, including transcription factors associated with embryogenesis, patterning, signaling, cell differentiation and neural functioning (voltage gated channels, neurotransmitter biosynthesis and vesicle transport systems) [Srivastava *et al.* 2008]. Due to these features, *T. adhaerens* represents a model organism for answering how cells lacking hierarchical organization coordinate motion and respond to environmental cues.

As a benthic motile animal that must forage for its algal and bacterial prey but lacks predefined directionality, its ability to sense environmental cues and establish a multicellular leading edge dynamically is paramount to survival. Collective cellular decision-making is thus directly related to this animal's foraging efficiency and therefore to its fitness. The natural size variation of these animals, from 100 to 2000 micrometers in diameter, and constantly fluctuating amoeboid shape allows us to test how migration accuracy and coordination of cell groups with minimal structure scales with group size, and facilitated the evolution of animal multicellularity. The natural size variation of *T. adhaerens* also enables the testing of theories on efficiency and scaling of information propagation in groups. It has recently been suggested, from principles in statistical mechanics, that collectively moving systems are self-organized critical systems in which information can propagate throughout an entire group irrespective of group size [Cavagna *et al.* 2010; Mora and Bialek 2011; Handegard *et al.* 2012]. Our organism enables the testing of this theory for the first time at multiple orders of magnitude.

Also notable is the fact that our experimental animal, which reproduces asexually by fission but never fuses, is ideal for investigating group size stability with fission-only groups, which has been largely-ignored in spite of substantial investigation in fission-fusion groups [Sibly 1983; for a review see Sumpter 2010].

## 2. EXPERIMENTAL INVESTIGATIONS AND METHODOLOGY

We will test two decision-making problems in this animal: firstly, how this animal establishes transient polarity in an otherwise isotropic body plan in response to chemotactic signals, and secondly, how this animal maintains an optimal size in different environments. We will identify how this animal tracks gradients and determine if directional changes in animal motion correspond to transience of leadership between different cells (sudden reversal of polarization), or due to changes in the position of leader cells (group rotation). We will also determine how gradient tracking accuracy and coordination scales with cellular group size. Once we identify an optimal size, we will investigate how this animal mediates its asexual reproduction through fission to optimize its size for different environments.

Long (2 hours or more) time series of animal movement were recorded using a 4x objective lens and brightfield illumination coupled to an automated stage operated through MATLAB. The thin body plan of *T. adhaerens*, consisting of only three cell layers, allows for us to use optical density as a proxy for nuclei (Fig. 1A), which we then use to track cellular movement by particle image velocimetry (PIV, Fig. 1B). Using this information, we can then determine the spatial length at which movements of different parts of the animal become uncorrelated (Fig. 1C), testing if there is a physical limit to coordination and information propagation. We can then relate the observed shape and internal dynamics of each animal to directional changes in an animal's movement over time (Fig. 2A) to the mechanism by which a leading edge is established. We further intend to quantify gradient tracking by *T. adhaerens*, be it chemotaxis on surfaces with pasted algal patterns, or negative phototaxis against harmful ultra-violet light. *T. adhaerens* is known to exhibit both substrate preference and behavioral/exploratory differences when exposed to different algal concentrations and UV light [Ueda *et al.* 1999; Pearse and Voigt 2007], but taxis and its scaling with animal size has never been tested.

## 3. PRELIMINARY RESULTS AND FUTURE WORK

The magnitude of internal velocity fluctuations in *T. adhaerens*, as measured by pair-wise correlations between different PIV-generated vectors, indicates that such fluctuation sizes increase linearly with animal size (Fig. 1D), a trend found in other systems believed to show scale-free information transfer. We expect larger *T. adharens* individuals that have a greater body length over which to integrate environmental gradients (Fig. 3A) will also have greater chemotactic accuracy (Fig. 3B), in accordance with theoretical predictions [Simons 2004] and empirical results in other systems [Berdahl et al. 2013]. Future work will focus on relating the our discoveries in the intercellular dynamics of *T. adhaerens* (Fig. 1B) to its exploratory behavior which we observed, especially its directional persistence over short time scales (Fig. 2B) and its burst-like locomotion (Fig. 2C). We will then test whether the chemotactic accuracy of *T. adhaerens* scales with size according to our predictions (Fig. 3) or if there is a coordination cost that inhibits accurate chemotaxis in the largest animals.

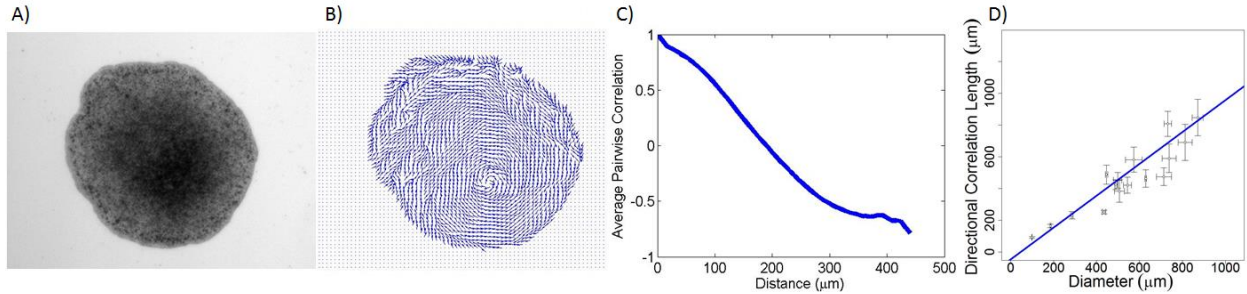

Fig. 1. A) The thin body plan of T. adhaerens allows for the use of optical density to track cellular positions within the animal B) Vector field generated by optical flow analysis describing flow of tissue and cells within the animal; C) Average pair-wise directional correlation between movement vectors at different separations. The correlation length is defined as the point of zero-crossing. D) Average correlation lengths of cellular movement at different organismal sizes.

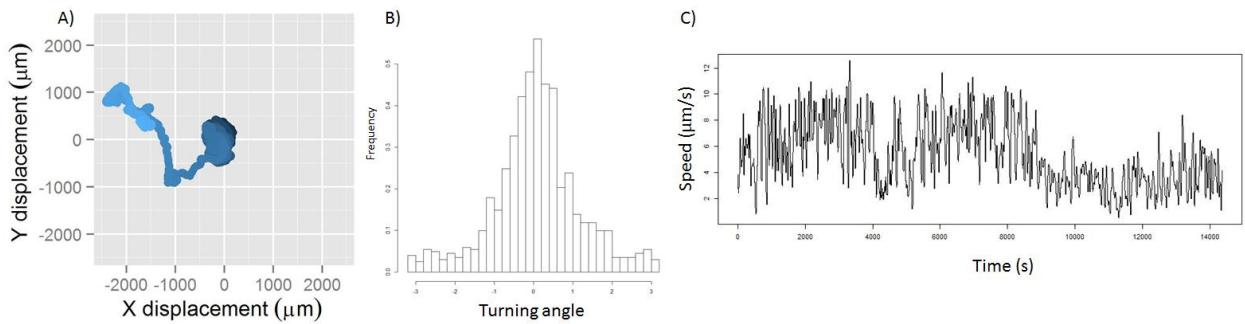

Figure 2: A) Trajectory of animal over a two hour recording. B) Distribution of turning angles between displacements separated by 15 s intervals, displaying directional persistence. C) Speed of animal over time shows periodic bursts of locomotion.

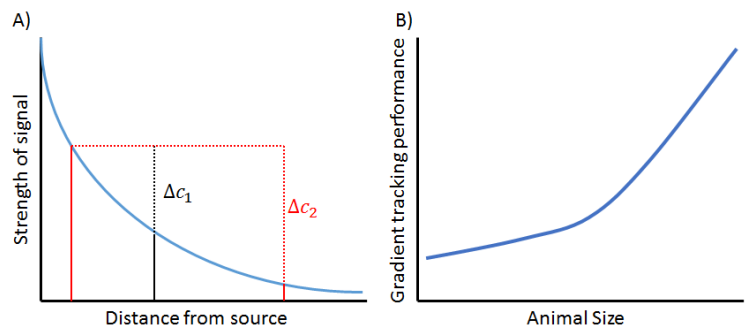

Fig. 3: A) Larger animals are able to integrate a directional gradient over a greater length, resulting in a greater concentration difference between the leading and trailing edges of the animal body. B) We expect animals of larger sizes to more accurately track gradients by exploiting the ability to integrate a chemical cue over a larger body diameter.